\documentclass[prd,floats,floatfix,showpacs,nofootinbib]{revtex4}
\usepackage{graphicx}
\usepackage{graphics}
\usepackage{amssymb}
\usepackage{amsmath}
\usepackage{url}
\usepackage{color}
\newcommand{\fracc}[2]{\frac{\textstyle{#1}}{\textstyle{#2}}}

\begin{document}

\title{Magnetic fields and the Weyl tensor in the early universe}

\author{E. Bittencourt$^{1}$}\email{eduardo.bittencourt@icranet.org}
\author{J. M. Salim$^2$}%\email{jsalim@cbpf.br}
\author{G. B. Santos$^{1}$}%\email{grasiele.dossantos@icranet.org}

%\affiliation{$^1$Funda\c c\~ao CAPES, Minist\'erio da Educa\c c\~ao, Bras\'ilia, Brasil}
\affiliation{$^1$Sapienza Universit\`a di Roma - Dipartimento di Fisica\\
P.le Aldo Moro 5 - 00185 Rome - Italy and\\
ICRANet, Piazza della Repubblica 10 - 65122 Pescara - Italy}
\affiliation{$^2$Instituto de Cosmologia Relatividade e Astrofisica ICRA - CBPF\\ Rua Doutor Xavier Sigaud, 150, CEP 22290-180, Rio de Janeiro,
Brazil}

\date{\today}

\begin{abstract}
We have solved the Einstein-Maxwell equations for a class of metrics with constant spatial curvature by considering only a primordial magnetic field as source. We assume a slight modification of the Tolman averaging relations so that the energy-momentum tensor of this field possesses an anisotropic pressure component. This inhomogeneous magnetic universe is isotropic and its time evolution is guided by the usual Friedmann equations. In the case of a flat universe, the space-time metric is free of singularities (except the well-known initial singularity at t = 0). It is shown that the anisotropic pressure of our model has a straightforward relation to the Weyl tensor. We then analyze the effect of this new ingredient on the motion of test particles and on the geodesic deviation of the cosmic fluid.
\end{abstract}

\maketitle

\section{Introduction}
The presence of magnetic fields in the early universe is fundamental to explain the evolution of the large scale structures of the universe. The current understanding of structure formation indicates that, beyond dark matter, magnetic fields are necessary to form galaxies and stars in accordance to what we observe (for a review see \cite{durrer} and references therein). In particular, magnetic fields are currently studied in the context of small perturbations of the standard cosmological model \cite{barrow,pogosian} with possible deviations caused by its presence \cite{caprini,shaw}. %However, if these magnetic fields are very intense in the standard cosmological model (SCM), they can modify considerably the nucleosynthesis (refs) and it is certainly an undesirable consequence.
On the other hand, aiming to produce a more realistic description of the universe, some authors claim that the main problem of the standard cosmological model (SCM) lies on the huge simplification of the Friedmann-Lema\^itre-Robertson-Walker (FLRW) geometry adopted as a background. Hence, in a variety of ways, they suggest modifications of spacetime symmetries leading to inhomogeneous cosmological models \cite{lemaitre,tolman,bondi,szekeres,mcvittie,wiltshire} (for a review see \cite{ltb}). In both cases (SCM and inhomogeneous models), magnetic fields are widely studied and should have interesting consequences, for example the avoidance of initial singularity \cite{delorenci02,aline}.

We consider in this work the evolution of a cosmological model in the presence of a primordial magnetic field. In addition to that, we slightly modify the Tolman relations for the averages of the electromagnetic field, and that renders our model inhomogeneous. The effective energy-momentum distribution of the magnetic field in this case possesses an anisotropic pressure term $\pi_{\mu\nu}$ that provides non-trivial properties for the metric.

As mentioned in Ref.\ \cite{ltb}, $\pi_{\mu\nu}$ has been indeed considered in the establishment of alternative models, but it is ruled out in standard cosmology due to the violation of the FLRW symmetries and its association restricted to the shear tensor as dictated by the fluid mechanics \cite{landau}. However, in this work we show that the Einstein-Maxwell equations allow a set of solutions which is driven by the usual Friedmann equations and has a 3-space of constant spatial curvature, given by the Schwarzschild-de Sitter lattice \cite{rindler}. The source of these equations may be represented by a radiation fluid ($p=\rho/3$) with an extra anisotropic pressure component. In contrast to most inhomogeneous cosmological models, ours is part of the shear-free solutions of Einstein equations \cite{ellis}. Therefore, $\pi^{\mu}{}_{\nu}$ should be linked to another symmetric traceless quantity which in this case it is shown to be the electric part of the Weyl tensor.

This work is organized as follows. In Sec.\ \ref{II} we introduce the model with the stochastic magnetic field as source for the Einstein equations and see the consequences of the presence of an anisotropic pressure component due to this field on the evolution equations for the scale factor. We then analyze qualitatively the case of a flat space and the relation between $\pi^{\mu}{}_{\nu}$ and the Weyl tensor. In Sec.\ \ref{III} we analyze the kinematical properties of the model through the study of geodesic motion and geodesic deviation. We finally present our conclusions in Sec.\ \ref{IV}.

\section{Magnetic field as a source of anisotropic pressure}\label{II}
We start from the linear Lagrangian of Maxwell's theory of electromagnetism
$$L=-\fracc{1}{4}\,F,$$
where $F\equiv\,F^{\mu\nu} F_{\mu\nu} = 2 (B^2-E^2)$.
The energy-momentum tensor corresponding to this Lagrangian is
$$T_{\mu\nu}=F_{\mu}{}^{\alpha}F_{\alpha\nu}-Lg_{\mu\nu},$$
whose decomposition into irreducible parts with respect to a normalized time-like vector field $V^{\mu}$
yields
\begin{equation}
\label{rho_p}
\rho = \frac{1}{2}(E^2+B^2), \hspace{.5cm} p =\frac{\rho}{3},
\end{equation}
\begin{equation}
\label{q}
q^{\alpha} = \eta^{\alpha}{}_{\beta\mu\nu} V^{\beta} E^{\mu} B^{\nu}
\end{equation}
and
\begin{equation}
\label{pi_em}
\pi_{\mu\nu} = -E_{\mu} E_{\nu} -B_{\mu} B_{\nu} - \frac{1}{3} (E^2+B^2) h_{\mu\nu},
\end{equation}
where $\rho$ is the energy density, $p$ is the isotropic pressure, $q^{\alpha}$ is the heat flux and $\pi_{\mu\nu}$ is the anisotropic pressure. We denote $h_{\mu\nu}\equiv g_{\mu\nu}-V_{\mu}V_{\nu}$ as the projector on the 3-space orthogonal to $V^{\mu}$. Greek indices run in the range $(0,1,2,3)$ and Latin indices
run in the range $(1,2,3)$.

The infinitesimal line element we use to describe the model is given by

\begin{equation}
\label{fried}
ds^2=dt^2-a^2(t)[d\chi^2+\sigma^2(\chi)d\Omega^2],
\end{equation}
where $t$ represents the cosmic time, $a(t)$ is the scale factor and $\sigma(\chi)$ is an arbitrary function of the spatial coordinate $\chi$.

A straightforward calculation yields the curvature of the spatial sections of constant time, namely
\begin{equation}
\label{3_ricci}
^{(3)}R=-2\left(2\fracc{\sigma''}{\sigma}+\fracc{\sigma'^2}{\sigma^2}-\fracc{1}{\sigma^2}\right),
\end{equation}
where prime ($\,'\,$) means derivative w.r.t. $\chi$. Once these spatial sections are isotropic, it is possible to assume that $^{(3)}R$ has the same value everywhere, i.e., $^{(3)}R\equiv6\epsilon$ where $\epsilon$ is a constant. Therefore, the scalar curvature

\begin{equation}
\label{4_ricci_simp}
R=6\left(\fracc{\ddot a}{a}+\fracc{\dot a^2}{a^2}+\fracc{\epsilon}{a^2}\right)
\end{equation}
becomes only a function of time, where dot ($\,\dot{} \,$) means time derivative.

Due to the special symmetries of the metric\ (\ref{fried}), the electromagnetic field can be considered as source of the gravitational field only if an averaging process is performed (cf.\ Ref.\ \cite{tolman2,hind}). The standard way to compute the volumetric spatial average of an arbitrary quantity X at the instant $t=t_0$ is defined by

\begin{equation}
\label{av_def}
\overline{X}\,\equiv \lim_{V \rightarrow V_0} \fracc{1}{V} \int X\sqrt{-g}d^3x^i,
\end{equation}
where we denote $V\equiv\int \sqrt{-g}d^3x^i$ and $V_0$ is a sufficiently large time dependent spatial volume (for details see \cite{gasperini}). Therefore, the mean values of the electric $E_i$ and magnetic $H_i$ fields are usually given by the so-called Tolman relations:

\begin{equation}
\label{rel_tol_1}
\overline{E_i}=0, \hspace{.3cm} \overline{H_i}=0, \hspace{.3cm} \overline{E_iH_j}=0,
\end{equation}

\begin{equation}
\label{rel_tol_2}
\overline{E^iE_j}=-\frac{1}{3}E^2h^i{}_j,
\end{equation}

\begin{equation}
\label{rel_tol_3}
\overline{B^iB_j}= -\frac{1}{3}B^2h^i{}_j.
\end{equation}

In this paper we are interested in the gravitational influence of the magnetic fields in the early universe. Thus, we set the electric field equal to zero---observational data indicate that magnetic fields play a more important role than electric fields at that epoch due to the high conductivity of the primordial plasma. Moreover, we do not impose homogeneity of the space-time but keep only the isotropy. Therefore, part of the Tolman relations are preserved [Eq.\ (\ref{rel_tol_1})] and the Maxwell equations are trivially verified by the mean fields. However, our choice leads us to slightly modify the Tolman relation concerning the second moment of the magnetic field (\ref{rel_tol_3}), as follows

\begin{equation}
\label{rel_tol_3_mod}
\overline{B^iB_j}=-\frac{1}{3}B^2h^i_j-\pi^{i}{}_{j},
\end{equation}
where we introduce an arbitrary traceless matrix $\pi^{i}{}_{j}$ that will be identified to an anisotropic pressure term and will have its components explicitly exhibited using Einstein equations in what follows.

Substitution of these hypotheses into Eqs.\ (\ref{rho_p})-(\ref{pi_em}) gives an effective energy-momentum distribution with respect to the comoving four-velocity $V^{\mu}=\delta^{\mu}_0$ described by a radiation fluid with energy density $\rho=B^2/2$ and isotropic pressure
$p=\rho/3$, and an anisotropic pressure term $\pi^{\mu}{}_{\nu}$. The manner in which we introduce such modification implies that it has necessarily the same symmetries of an anisotropic pressure term, that is, it is symmetric and traceless.

The next step is to encounter any possible expression for $\pi^{\mu}{}_{\nu}$ compatible with the metric we have set\footnote{Although it may seem that the procedure that follows is just an {\em Ansatz}, the expressions found are physically justified by the relation between $\pi^{\mu}{}_{\nu}$ and the Weyl tensor which will be given afterwards.}. We calculate the non-trivial components of the Einstein equations, $G^{\mu}{}_{\nu}=-T^{\mu}{}_{\nu}$, where the Einstein constant is set to $1$, and that results

\begin{subequations}
\label{eins_tens_pi}
\begin{eqnarray}
&&3\fracc{\dot a^2}{a^2}+3\fracc{\epsilon}{a^2}=\rho,\label{eins_tens0_pi}\\[2ex]
&&2\fracc{\ddot a}{a} + \fracc{\dot a^2}{a^2} + \fracc{3\epsilon}{a^2} + \fracc{2}{a^2}\fracc{\sigma''}{\sigma} = -p+\pi^1{}_1,\label{eins_tens1_pi}\\[2ex]
&&2\fracc{\ddot a}{a} + \fracc{\dot a^2}{a^2} - \fracc{1}{a^2}\fracc{\sigma''}{\sigma} = -p+\pi^2{}_2.\label{eins_tens2_pi}
\end{eqnarray}
\end{subequations}
The off-diagonal components are identically zero.

From Eqs.\ (\ref{eins_tens_pi}) we can see that the Einstein equations admit a solution with constant spatial curvature and an anisotropic pressure term only if $\pi^{\mu}{}_{\nu}$ is given by
\begin{equation}
\label{sol_pi}
\pi^{2}{}_{2} = \pi^{3}{}_{3}, \hspace{1cm} \pi^{1}{}_{1} = -2\pi^{2}{}_{2}, \hspace{.5cm} \mbox{and} \hspace{.5cm} \pi^{1}{}_{1}=\fracc{2k}{a^2\,\sigma^3},
\end{equation}
where $k$ is an integration constant. Note that the Einstein equations impose that $\pi^i{}_j=0$ for $i\neq j$. Moreover, the constancy of the tri-curvature imposes that $\sigma(\chi)$ must satisfy the following first-order differential equation

\begin{equation}
\label{dif_sigma}
\sigma'=\pm\sqrt{1-\epsilon \sigma^2-\fracc{2k}{\sigma}}.
\end{equation}

It should be remarked that the presence of the anisotropic pressure does not affect the time evolution of the scale factor, which is exactly guided by the Friedmann equations

\begin{subequations}
\label{gr_fried_solved}
\begin{eqnarray}
&&\left( \fracc{\dot a}{a} \right)^2 = \fracc{\rho}{3} - \fracc{\epsilon}{a^2},\label{gr_fried_solved1}\\[2ex]
&&\fracc{\ddot a}{a}=-\fracc{1}{6}(1+3\lambda)\rho,\label{gr_fried_solved2}
\end{eqnarray}
\end{subequations}
where $\lambda$ is the equation of state parameter which in our case reads $\lambda=1/3$.
Solving these equations for the magnetic field, we find that its modulus is only a function of time given by
$$B=\fracc{B_0}{a^2},$$
where $B_0$ is a constant.

By substituting these results into Eq.\ (\ref{rel_tol_3_mod}) we can calculate the second moment of the magnetic field allowed by the Einstein equations. Then, we find out that the inhomogeneity of the space-time appears only in the second order of the average process, but it is not negligible, as we shall see in the analysis of a test particle motion afterwards.

\subsection{Qualitative analysis of $\sigma(\chi)$ and the special case of a flat universe}
In the general case Eq.\ (\ref{dif_sigma}) cannot be analytically integrated. However, some information can be obtained according to the qualitative theory of dynamical systems \cite{smale}. Depending on the value of $\epsilon$ and $k$, the polynomial inside the square root of Eq.\ (\ref{dif_sigma}) has different numbers of real roots. Thus, the analysis of the discriminant of the third-order polynomial $\Delta=4\epsilon(1-27\epsilon k^2)$ yields

\begin{equation}
\label{discrim}
\left\{\begin{array}{lcl}
\Delta>0&\Longrightarrow&\mbox{3 distinct real roots},\\[2ex]
\Delta=0&\Longrightarrow&\mbox{3 real roots, one being degenerate},\\[2ex]
\Delta<0&\Longrightarrow&\mbox{1 real root and 2 complex conjugates.}
\end{array}\right.
\end{equation}
This result implies that $k$ behaves as a bifurcation parameter breaking the topological symmetry of the possible spatial hypersurfaces in the FLRW metric for each value of $\epsilon$, once the sign of the latter is not enough to determine the sign of $\Delta$, which now depends also on the specific value of $k$. The key point we pick up from this simple analysis is that, fixing $\epsilon$, the value of $k$ is important in the determination of the range of $\sigma(\chi)$ and, in its turn, the spatial features of the metric.

\begin{figure}[!htb]
\centering
\includegraphics[width=6cm,height=8cm,angle=-90]{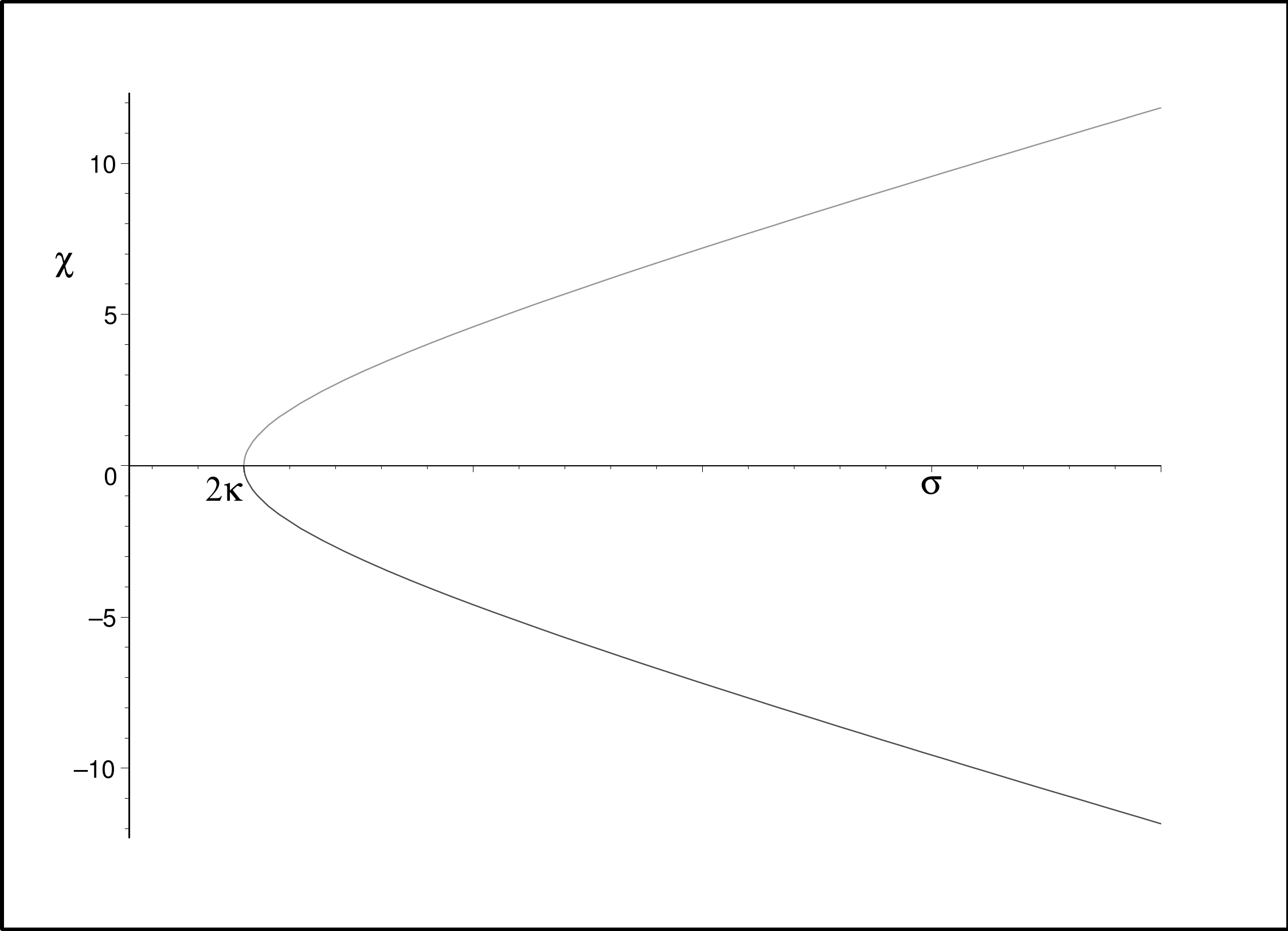}
\caption{Implicit plot of $\sigma(\chi)$ given by Eq.\ (\ref{sol_sigma}) when $\epsilon=0$. Note that $-\infty<\chi<\infty$, while $\sigma$ has a minimal value at $2k$. For large values of $|\chi|$, $\sigma(\chi)$ tends to the FLRW solution $\sigma(\chi)=\chi$.}
\label{fig1}
\end{figure}

The particular case where the tri-curvature is flat agrees with observational data. Thus, we carry on a more detailed analysis of this case, which is also simpler. For $\epsilon=0$, Eq.\ (\ref{dif_sigma}) can be analytically integrated resulting in

\begin{equation}
\label{sol_sigma}
\chi = \pm\left[ \sqrt{\sigma^2-2k\sigma} + k\ln\left(\frac{\sigma-k+\sqrt{\sigma^2-2k\sigma} }{k}\right)\right],
\end{equation}
which gives $\sigma$ implicitly in terms of $\chi$ as illustrated in Fig.\ (\ref{fig1}).
We are restricted to the case $k>0$ because if we let $k$ assume the opposite sign, then the metric tensor would possess a naked singularity at $\sigma=0$, violating the cosmic censorship \cite{penrose} if it holds true. Thus, there is a minimum value for $\sigma$ at $2k$ when $\chi=0$. It means that the region $\sigma<2k$ is excluded from the manifold and this statement is corroborated by Eq.\ (\ref{dif_sigma}) once $\sigma'$ becomes complex in this hypothetical region. Due to the maximal extension of the range of each spatial coordinate, we can assert that the manifold is completely covered. These results will be very important in the analysis of the geodesic motion and in the demonstration of the geodesic completeness and smoothness of our solution\footnote{We have also calculated all the algebraic curvature invariants and verified their regularity at $\sigma=2k$, see Appendix for details.}, at least for $\epsilon=0$.

\subsection{The non-vanishing Weyl tensor}
Making the coordinate transformation $r=\sigma(\chi)$, the line element given by Eq.\ (\ref{fried}) becomes

\begin{equation}
\label{fried_r_pi}
ds^2=dt^2-a^2(t)\left(\fracc{dr^2}{1-\epsilon r^2-\frac{2k}{r}}+r^2d\Omega^2\right).
\end{equation}
Note that the time evolution of this metric is still given by the usual Friedmann equations\ (\ref{gr_fried_solved1}) and\ (\ref{gr_fried_solved2}). From this line element, the physical meaning of the parameter $k$ is enlightened: it is related to the scale of homogeneity of the universe; if $2k/r\ll1$ this term can be dropped out in Eq.\ (\ref{fried_r_pi}) and, hence, the Friedmann model is completely recovered in this regime; on the other hand, if $2k/r$ is not negligible, the 3-geometry is non-trivial and the particle trajectories in this metric have special features that we shall discuss in the next sections. It should also be remarked that this solution is not conformally equivalent to Schwarzschild-de Sitter metric, although they have the same lattice surface (see Ref.\ \cite{rindler} for more details). From this similitude, the analysis of the Killing vectors and the Petrov classification is straightforward and we conclude that the metric\ (\ref{fried_r_pi}) has only 3 Killing vectors, corresponding to the spacetime isotropy and, consequently, it is Petrov-type D.

Going back to the anisotropic pressure given in Eq.\ (\ref{sol_pi}), we note that $\pi^{\mu}{}_{\nu}$ has a spatial dependence that has the same power as the Newtonian tidal forces ($\sim r^{-3}$). Still more impressive is that Eq.\ (\ref{sol_pi}) is the unique solution compatible with the Einstein equations in this model. This raises the question: is such expression a mere coincidence or does it come from a more fundamental relation between the gravitational field and its geometrization procedure? This question cannot be answered satisfactorily in the Einstein approach because the main part of the curvature tensor responsible for the tidal forces (the Weyl tensor) is absent in the Einstein equations. Therefore, we have to appeal to the Bianchi identities written in terms of the Weyl tensor--which are also known as the quasi-Maxwellian equations of general relativity \cite{ellis_qm}--in order to explain such relation.

From the time evolution of the shear tensor, a straightforward calculation yields the electric part of Weyl tensor for the observer's four-velocity $V^{\mu}=\delta^{\mu}_0$ as follows

\begin{equation}
\label{wyel}
E_{\mu\nu}=-\fracc{1}{2}\pi_{\mu\nu}.
\end{equation}
%or, in components, it is explicitly given by

%\begin{equation}
%\label{el_weyl_r}
%E^{2}{}_{2} = E^{3}{}_{3}, \hspace{.3cm} %E^{1}{}_{1} = -2E^{2}{}_{2}, \hspace{.3cm} \mbox{and} \hspace{.3cm} E^{1}{}_{1}=-\fracc{k}{a^2r^3}.
%\end{equation}

We see that, even assuming a constant spatial curvature, the Weyl tensor does not vanish and is given by the anisotropic pressure, which behaves as Newtonian tidal forces (multiplied by a time dependent function). Therefore, we are led to conclude that the Friedmann equations in the presence of an anisotropic pressure modify the Weyl tensor via scale factor, but the presence of the Weyl tensor does not change the time evolution of the spacetime.

\section{Kinematical analysis}\label{III}

Let us now analyze the consequences of the non-null Weyl tensor on the trajectories of test particles and on the geodesic deviation. It is shown that the test particle motion should be different from the one provided by FLRW metrics. However, we shall see that the geodesic deviation is unchanged and that renders the effects of this new component of the model unobservable at the non-perturbative level.

\subsection{Trajectories of test particles}\label{test_part}

Due to the isotropy of the spacetime we are dealing with, instead of integrating completely the geodesic equations of the metric\ (\ref{fried_r_pi}), the analysis of the test particle trajectories (time-like and light-like) performed in the equatorial plane ($\theta=\pi/2$ and $\dot\theta=0$) is enough to characterize all we need to know about the geodesic motion. This simplification allows us to study the behavior of the effective potential to which these particles are subjected and to compare their paths with other cases. Thus, the simplified geodesic equations reduce to

\begin{eqnarray}
\label{geo_pi}
&&t''+\fracc{a\dot a}{A}r'^2+a\dot ar^2\phi'^2=0,\label{geo_pi1}\\[2ex]
&&t'^2-\fracc{a^2}{A}r'^2-a^2r^2\phi'^2=b,\label{geo_pi2}\\[2ex]
&&\phi''+2\fracc{\dot a}{a}t'\phi'+2\fracc{r'}{r}\phi'=0,\label{geo_pi3}
\end{eqnarray}
where we denoted $X'\equiv dX/d\tau$, $\tau$ is the affine parameter along the curve and $b$ is either $1$ or $0$ for timelike and light-like geodesics, respectively. We also denoted $A(r)\equiv1-\epsilon\, r^2-2k/r$.

Firstly, we solve Eq.\ (\ref{geo_pi3}) and get the angular momentum conservation $\phi'=l/a^2r^2$, where $l$ is an integration constant. Then, substituting this equation and Eq.\ (\ref{geo_pi2}) into Eq.\ (\ref{geo_pi1}) we obtain the equation $t'^2-b=E/a^2$, where $E$ is another integration constant. Substituting these equations in Eq.\ (\ref{geo_pi2}) yields

\begin{equation}
\label{int_geo_pi_r}
a^4r'^2=\left(E-\frac{l^2}{r^2}\right)A.
\end{equation}
This equation fixes $E>0$ and can be seen as the energy conservation equation of a particle moving in a one-dimensional effective potential. Note that it has the constant ``mechanical energy" $E$, the kinetic-like term $a^4r'^2$ and the remaining ones correspond to an effective potential which will be denoted by $V(r)$.

\begin{figure}[!htb]
\centering
\includegraphics[width=9cm,height=6cm]{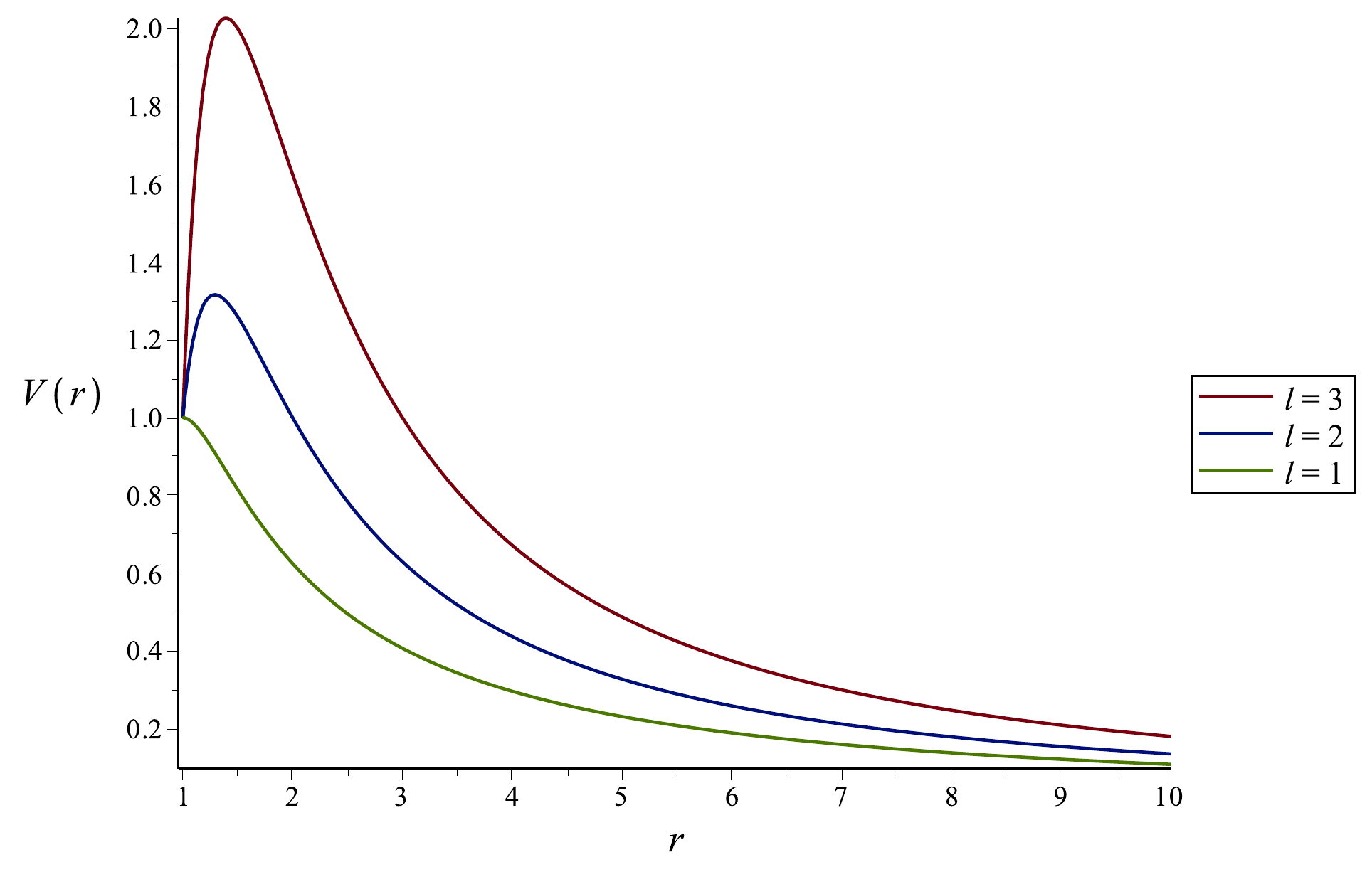}
\caption{Effective potential for different values of the angular momentum, with $E=1$ and $k=0.5$. Note that in this case the coordinate $r$ attains its minimum value at $r=1$.}
\label{fig2}
\end{figure}

In order to compare this case with the Schwarzschild one, we set $\epsilon=0$ in Eq.\ (\ref{int_geo_pi_r}). Therefore, $V(r)$ is given by

%\begin{equation}
%\label{int_geo_pi_r_e0}
%a^4r'^2=E-\fracc{2kE}{r}-\frac{l^2}{r^2}+\frac{2kl^2}{r^3}
%\end{equation}
%and, in this case, $V(r)$ is given by

$$V(r)=\fracc{2kE}{r}+\frac{l^2}{r^2}-\frac{2k\,l^2}{r^3}.$$
This potential possesses almost the same terms as the ones provided by the Schwarzschild metric for a single particle moving along geodesics, except the Newtonian one ($\sim r^{-1}$) which has a positive sign ($k,E>0$). Its behavior for different values of $l$ can be seen in Fig.\ (\ref{fig2}). As a consequence, for particles moving radially ($l=0$) or for large values of $r$, the effective gravitational potential is repulsive. It should also be remarked that when $r=2k$ the right hand side of Eq.\ (\ref{int_geo_pi_r}) is identically zero for massive particles and light rays. It means that the total energy is equal to the effective potential and, contrary to the Schwarzschild geodesics, this is a turning point for all test bodies. This lead us to conclude by other means that the region $r<2k$ is excluded from the manifold.

For the sake of comparison, we analyze qualitatively the Kepler problem in this solution rewriting Eq.\ (\ref{int_geo_pi_r}) in terms of the variable $u=1/r$ and seeking for the planetary orbits $u=u(\phi)$. For practical reasons, we consider the scale factor almost constant and study the test particle behavior in an interval of the cosmological time $(t_0,t_1)$, i.e., $a(t)\equiv a_0$ for $t\in(t_0,t_1)$. In other words, we assume that the cosmological evolution is very slow when compared to the period of revolution around the center of symmetry of our solution. We thus proceed similarly to the Schwarzschild case and we get the following equation for $u=u(\phi)$:

\begin{equation}
\label{diff_geo_pi_u}
\fracc{d^2u}{d\phi^2}+u=3ku^2-\fracc{Ek}{l^2}.
\end{equation}
This differential equation is very similar to the one given by the Schwarzschild metric in response to the Kepler problem if we interpret $k$ as the effective mass of the gravitational source and $E$ as the total mass of the test particle \cite{adler}. However, the inhomogeneous term of the equation has a negative sign. In qualitative terms, Eq.\ (\ref{diff_geo_pi_u}) can reproduce perihelion shifts for massive test particles ($b>0$) and light rays deflection ($b=0$), whereas the intensities of these phenomena can be provided only through the complete integration of that equation.

These results indicate how the presence of the Weyl tensor (or, equivalently, $\pi_{\mu\nu}$) affects the geodesic motion of test particles. Note that in order to determine the local gravitational effects that cannot be interpreted as being induced by any local ``visible" matter distribution, the Weyl tensor must be considered as it contains information of the global conditions imposed upon the spacetime, which in this case are related to the configuration of a stochastic magnetic field on this spacetime.

\subsection{Geodesic deviation of the cosmological fluid}

Among its attributes, the quasi-Maxwellian representation of gravity has the quality of putting together the formalism of the electromagnetic interactions and a formal approach to general relativity. Nonetheless, some fundamental distinctions must be stressed. The empirical determination of an electromagnetic field, for instance, is made through the Lorentz force and a test particle in order to identify the presence of the electromagnetic field. The electromagnetic tensor, obtained from the integration of the Maxwell equations, does not distinguish the contribution of local charge and current distributions from boundary conditions. In general relativity, the empirical identification of a gravitational field cannot be made using a single test particle since the Christoffel symbols can be set equal to zero by coordinate transformations. So, in order to empirically determine the properties of the gravitational field, it is necessary to look for the geodetic deviation expressed in terms of the curvature tensor. This tensor separates explicitly the contribution coming from the local distribution of the energy-momentum tensor, algebraically associated to the traces of the Riemann tensor, from the global contribution of boundary conditions represented by the Weyl tensor.

Indeed, the measurements of the gravitational field effects can only be done through the geodesic deviation equation which determines the rate of the relative acceleration between two infinitesimally nearby geodesics, namely

\begin{equation}
\label{geo_dev}
\fracc{d^2z^{\alpha}}{ds^2}=-R^{\alpha}{}_{\beta\mu\nu}V^{\beta}z^{\mu}V^{\nu},
\end{equation}
where $z^{\alpha}$ is the deviation vector and $V^{\mu}$ is the vector field tangent to the geodesic congruence.

The distortion produced by the Weyl tensor upon a given congruence of curves can be accounted for through the substitution of the Riemann tensor by its decomposition into irreducible parts. Since we are dealing with the comoving frame of the cosmological fluid, we set $V^{\mu}=\delta^{\mu}_0$. Evaluating the right hand side of Eq.\ (\ref{geo_dev}), we get
$$\fracc{d^2z^{\alpha}}{ds^2} = \left(E^{\alpha}{}_{\mu} + \frac{1}{2}\pi^{\alpha}{}_{\mu}\right) z^{\mu}+\fracc{1}{6}(\rho+3p)h^{\alpha}{}_{\mu}z^{\mu}.$$
According to Eq.\ (\ref{wyel}), the remarkable result coming from the expression above is that the term inside the big brackets is identically zero for the metric\ (\ref{fried_r_pi}). Therefore, the cosmological fluid does not feel the effects due to the presence of the anisotropic pressure and the Weyl tensor. In other words, the distortion caused by the anisotropic pressure and the electric part of the Weyl tensor are compensated in such a way that the cosmological observers do not attribute any eventual modification of the spacetime to these quantities, enabling one to set $E_{\mu\nu}$ and $\pi_{\mu\nu}$ equal to zero {\it a priori}, erroneously. However, this is not allowed if we want to understand correctly the gravitational field effects in the Universe using the available empirical data as initial conditions. Besides, the presence of the anisotropic pressure may change dramatically the perturbed version of the theory and hence the large scale structure formation. This will be investigated in a forthcoming paper.

\section{Concluding Remarks}\label{IV}

From an exact solution including a primordial magnetic field we have seen that Einstein equations do not contain in its dynamics all the necessary information to determine the curvature tensor from empirical data. Namely, boundary conditions including the Weyl tensor, defined on a spatial hypersurface, cannot be provided by observational data, which lies in the past null cone of the observer. In other words, the Einstein equations should represent the universe as an open system rather than a closed totality, since at any moment new elements not included in the dynamical equations can play a role and modify the spacetime properties we observe. In particular, the presence of the Weyl tensor in this case shall change the perturbative version of the model and consequently the structure formation even if it cannot be perceived through the geodesic deviation.

The attempt in developing such cosmological model has also some features that should be explored further. For instance, the cases for which the tri-curvature is nonnull, which gives $k$ the status of a bifurcation parameter, should be addressed. Also, if we fix the tri-curvature to be flat but instead let $k$ assume the opposite sign we have considered in this paper, thus violating the cosmic censorship, then we get the correct Newtonian term in the effective potential for test particles at small values of the radial coordinate. For future developments, we also intend to investigate the relation between the magnetic field and the Weyl tensor interpreted as tidal forces from first principles using kinetic theory.

In summary, we have shown how the standard cosmological model sets the Weyl tensor equal to zero \textit{ab initio} and that this is not a consequence of the Friedmann equations. This was shown through the development of a cosmological model with constant spatial curvature and a non-zero Weyl tensor without spoiling the conventional time evolution of the universe through the Friedmann equations.

\begin{acknowledgements}
We would like to thank Prof. M. Novello for his comments on a previous version of this paper. EB and GBS are financially supported by the CAPES-ICRANet program through the grants BEX 13956/13-2 and 13955/13-6.
\end{acknowledgements}

\appendix

\section{Curvature Invariants}

Following the procedure in \cite{zakh}, we display here all the algebraic curvature invariants associated to the metric (\ref{fried_r_pi}) in order to guarantee the regularity of the solution in the case $k>0$ and its singular behavior when $k<0$.

The expressions for these invariants are:

\begin{equation}
\label{inv_def}
\begin{array}{lcl}
I_{0} &=& R,\\[2ex]
I_{1} &=& \frac{1}{4} S_{\mu\nu} S^{\mu\nu},\\[2ex]
I_{2} &=& -\frac{1}{8} S_{\alpha\beta} S^{\beta\mu} S_{\mu}{}^{\alpha},\\[2ex]
I_{3} &=& \frac{1}{16} S_{\alpha\beta} S^{\beta\mu} S_{\mu\lambda} S^{\alpha\lambda},\\[2ex]
I_{4} &=& \frac{1}{8} (W_{\alpha\beta\mu\nu} W^{\alpha\beta\mu\nu} +i\, W^{\alpha\beta\mu\nu}\,{}^{*}W_{\alpha\beta\mu\nu}),\\[2ex]
I_{5} &=& -\frac{1}{16} (W_{\alpha\beta}{}^{\rho\sigma} W_{\rho\sigma}{}^{\mu\nu}W_{\mu\nu}{}^{\alpha\beta} +i\, W^{\alpha\beta\rho\sigma} W_{\rho\sigma}{}^{\mu\nu}\,{}^{*}W_{\mu\nu\alpha\beta}),\\[2ex]
I_{6} &=& -\frac{1}{8} (S_{\mu\nu} D^{\mu\nu} +i\, S^{\mu\nu} \tilde{D}_{\mu\nu})\\[2ex]
I_{7} &=& \frac{1}{16}(D_{\mu\nu} D^{\mu\nu} - \tilde{D}_{\mu\nu} \tilde{D}^{\mu\nu}+2i\,\tilde{D}_{\mu\nu} D^{\mu\nu}),\\[2ex]
I_{8} &=& \frac{1}{16}(D_{\mu\nu} D^{\mu\nu} + \tilde{D}_{\mu\nu} \tilde{D}^{\mu\nu}),\\[2ex]
I_{9} &=& -\frac{1}{32}S^{\alpha\beta}(D_{\alpha\gamma}D^{\gamma}{}_{\beta}+\tilde D_{\alpha\gamma}\tilde D^{\gamma}{}_{\beta}),\\[2ex]
I_{10} &=& \frac{1}{32}[W^{\alpha\rho\sigma\beta}(D_{\alpha\beta} D_{\rho\sigma} - \tilde{D}_{\alpha\beta} \tilde{D}_{\rho\sigma}+i\, {}^{*}W^{\alpha\rho\sigma\beta}(D_{\alpha\beta} D_{\rho\sigma} - \tilde{D}_{\alpha\beta} \tilde{D}_{\rho\sigma})],\\[2ex]
I_{11} &=& \frac{1}{32} S_{\alpha}{}^{\mu} S_{\mu}{}^{\gamma} S_{\beta}{}^{\nu} S_{\nu}{}^{\delta} ( W^{\alpha\beta}{}_{\gamma\delta} +i\, {}^{*}W^{\alpha\beta}{}_{\gamma\delta}).
\end{array}
\end{equation}
where we denote $S_{\mu\nu} \equiv R_{\mu\nu} - (R/4) g_{\mu\nu}$, $D_{\mu\nu} \equiv W_{\mu\alpha\nu\beta} S^{\alpha\beta}$ and $ \tilde{D}_{\mu\nu} \equiv {}^{*}W_{\mu\alpha\nu\beta} S^{\alpha\beta}$. Solving the Friedmann equations for the energy-momentum tensor we have considered in the text (which yields $a(t)=\sqrt{t}$) and using the relation between the electric part of the Weyl tensor and the anisotropic pressure, we obtain the following non-null invariants

\begin{equation}
\label{inv_def_expl}
\begin{array}{ccl}
I_1 &=& \frac{3}{16}\frac{r^6+8k^2t^2}{t^4r^6},\\[2ex]
I_2 &=& -\frac{3}{64}\frac{r^9+16k^3t^3-12t^2k^2r^3}{t^6r^9},\\[2ex]
I_3 &=& \frac{3}{1024}\frac{7r^{12}+384k^4t^4-128t^3k^3r^3+48t^2k^2r^6}{t^8r^{12}},\\[2ex]
\mbox{Re}(I_4) &=& \frac{3}{2}\frac{k^2}{t^2r^6},\\[2ex]
\mbox{Re}(I_5) &=& \frac{3}{4}\frac{k^3}{t^3r^9},\\[2ex]
\mbox{Re}(I_6) &=& \frac{3k^2}{4}\frac{r^3+kt}{t^4r^9},\\[2ex]
\mbox{Re}(I_7) &=& I_8 = \frac{3k^2}{32}\frac{r^6+12k^2t^2+tkr^3}{t^6r^{12}},\\[2ex]
I_9 &=& \frac{3k^2}{256}\frac{r^9+16k^3t^3-12t^2k^2r^3}{t^8r^{15}},\\[2ex]
\mbox{Re}(I_{10}) &=& \frac{3k^2}{64}\frac{r^6+20k^2t^2+12tkr^3}{t^7r^{15}},\\[2ex]
\mbox{Re}(I_{11}) &=& \frac{3k^2}{64}\frac{r^9+4k^3t^3-3tkr^6}{t^8r^{15}}.
\end{array}
\end{equation}
Finally, it is straightforward to see that when $k>0$, which implies $r\in(2k,\infty)$, all the invariants are finite everywhere and in particular at $r=2k$. Otherwise, when $k<0$ and consequently $r\in(0,\infty)$, the invariants diverge at $r=0$ exhibiting a real singularity. The singularity at $t=0$ still remains in both cases.

\end{document}